\documentclass[11pt]{article}
\usepackage{amsmath}
\usepackage{amsfonts}
\usepackage{amssymb}
\usepackage{graphicx}
\usepackage[latin1]{inputenc}
\newcommand\exval[1]{\left\langle #1 \right\rangle}
\newcommand\ob{b}
\newcommand\bd{{b^\dagger}}
\newcommand\G{G}
\newcommand\nn{\nonumber}
\newcommand\op{}
\newcommand{\linsig}{linear-$\sigma$}
\renewcommand\epsilon{\varepsilon}
\newcommand\beq{\begin{equation}}
\newcommand\eeq{\end{equation}}
\begin{document}
 \title{ Some applications of renormalized RPA in bosonic field theories}
\author{H. Hansen$^{1,2,}$\footnote{E-mail: {\tt hansen@teor.fis.uc.pt}}, G. Chanfray$^{1}$ \\   
{\small\it $^{(1)}$ IPN Lyon,  43 Bd.  du 11 Novembre  1918, F-69622 Villeurbanne Cedex} \\ 
{\small\it $^{(2)}$ Centro de Física Teórica, Departamento de Física
da Universidade de Coimbra,}\\ 
{\small\it 3004-516 Coimbra, Portugal}}

\maketitle 
\begin{abstract} 
 
We present some applications of the renormalized RPA in bosonic field theories. 
We first present some developments for the explicit calculation of the total energy 
in $\Phi^4$ theory and discuss its phase structure in $1+1$ dimensions. We also demonstrate 
that the Goldstone theorem is satisfied in the $O(N)$ model within the renormalized RPA. 
 
\end{abstract}

\section{Introduction}
The application to quantum field theories of non-perturbative methods used in
the nuclear many-body problem \cite{RS80} has recently given rise to numerous promising works. 
One central motivation is to obtain tools to describe matter made of
strongly interacting hadrons in presence of broken symmetries such as chiral
symmetry. In particular the well-known  RPA method, which has been originally 
developed in the context of condensed matter physics, has been recently 
applied  to study a bosonic  $O(N)$ model ({\it i.e.} the linear sigma model). 
It has been demonstrated \cite{ACSW} that the standard RPA is able to restore the Goldstone 
theorem which is violated at the level of the usual variational Gaussian 
approximation \cite{S84,KV89,DMS96}. Although this result can be seen as a major success, the RPA method 
in its standard  form possesses some weak points. In particular it has the 
tendency to overestimate the attractive correlation energy, at
least in examples of nuclear physics. In a recent paper \cite{HCDS02}, hereafter referred as I, 
we have developed the formalism of a superior version of the RPA, namely the
renormalized RPA (r-RPA) in the particular context of $\lambda\Phi^4$ theory. 
As shown in I, one important merit of the r-RPA is to cure the instability problem
appearing in the standard RPA in $1+1$ dimensions. The purpose of this paper is to 
present a more detailed study of the phase structure of this theory within the 
r-RPA approach. Besides we introduce improvements of the
former calculation and obtain a second order phase transition for this model.
We also present the r-RPA  method for the $O(N)$ model,  demonstrating that the 
Goldstone theorem  (massless pions) is also satisfied at this level. Since all 
the details of the approach have been given in I, we limit ourselves to the strict 
minimum for what concerns the formalism.

\section{The  r-RPA in $\boldsymbol\Phi^4$ theory}

We consider the Lagrangian density: 
\begin{equation}
{\cal L}={1\over 2}\,\partial^\mu\Phi(x,t)\, \partial_\mu\Phi(x,t)\,-
{1\over 2}\,\mu_0^2\,\Phi^2(x,t)\,-\,{b\over 24}\Phi^4(x,t)
\end{equation}
where $\mu_0^2$ is a constant and the bare coupling constant $b=\lambda/6$
is positive for reasons of stability. We decompose the scalar field 
$\Phi(x,t)$ in a classical part
or condensate $s$ and a fluctuating piece $\phi(x,t)$:
\begin{equation}
\Phi(x,t)=\phi(x,t)\,+\,s,\qquad s=\langle\Phi(x,t)\rangle~.
\end{equation}
The presence of the condensate $s$ indicates a spontaneous breaking  of the
underlying $\Phi\to -\Phi$ symmetry. Introducing the conjugate field
$\Pi(x)$, one obtains for the Hamiltonian (in $d+1$ dimensions): 
\begin{eqnarray}
H = & &\int d^d x\,\,\bigg\{{1\over 2} \mu^2_0\,s^2\,+\,{b\over 24}\,s^4\,+\, 
\left(\mu_0\,s\,+\,{b\over 6}\,s^3\right)\,\phi(x)\,\nonumber\\
 & & + \,{1\over 2}\big[\Pi^2(x)\,+\,\left(\partial_i\phi\right)^2(x) \,+\,\big(
\mu^2_0\,+{b\over 2}\,s^2\big)\,\phi^2(x)\big]\nonumber\\
& & +\,{b\,s\over 6}\,\phi^3(x)\,+\,{b\over 24}\,\phi^4(x)\bigg\}\label{HAM0}~.
\end{eqnarray}
Putting the system in a large box of volume $V=L^d$, 
it is convenient to work in momentum space and to expand the fields according to:
\begin{equation}
\phi(x)={1\over \sqrt{V}}\,\sum_{\vec q}\,e^{i\vec q\cdot \vec x}\,\phi_{\vec
q}(t)\quad,
\qquad
\Pi(x)=-\,{i\over \sqrt{V}}\,\sum_{\vec q}\,e^{i\vec q\cdot \vec x}\,\Pi_{\vec q}(t)~,
\end{equation}
{\it i.e.}, in terms of creation and annihilation operators obeying
the standard canonical commutation relations:
\begin{equation}
\phi_{{\vec q}}=\sqrt{1\over 2 \kappa_{\vec q}}\,\left(b_{\vec q}\,+\,
b^\dagger_{-{\vec q}}\right),\qquad
\Pi_{{\vec q}}=\sqrt{\kappa_{\vec q}\over 2}\,\left(b_{\vec q}\,-\,
b^\dagger_{-{\vec q}}\right)~.\label{FIPI}
\end{equation}
The choice of the basis ({\it i.e.} the choice of the $\kappa_{\vec q}$) will come out 
as a part of the RPA solution. 

\bigskip\noindent
In I \cite{HCDS02} we have explicitly solved the r-RPA problem, using the Green's function method, 
taking into account one-particle $b^\dagger_{\vec q}$ and 
two-particle $b^\dagger_{\vec q}\, b^\dagger_{\vec q\, '},  
b^\dagger_{\vec q}\, b_{-\vec q\, '}$  excitation operators. We refer the 
reader to I for the detailed derivation and we  only quote here the main results 
(we also give in the appendix the results  for the 1-1 1-2 and 2-2 
Green's functions).

\bigskip\noindent
It is convenient to choose the basis 
$\kappa_{\vec q}=\varepsilon_{\vec q}$ where $\varepsilon_{\vec q}$ is  the 
generalized mean-field energy, solution of the gap equation
$\langle[H,b_{\vec q} \, b_{-{\vec q}}]\rangle=0$. This gap equation reads: 
\begin{equation}
\varepsilon_{\vec q}^2={\vec q\,}^2\,+\mu_0^2\,+\,{b\over 2} s^2\,+
{b\over 2}\langle\phi^2\rangle_R\equiv {\vec q\,}^2\,+\,m^2\label{GGAP}
\end{equation}
where $\langle\phi^2\rangle_R=(1/V)\sum_{\vec q}\,
\langle 1\,+\,2 b^\dagger_{\vec q}\, b_{\vec q}\rangle/2\varepsilon_{\vec q}$
is the self-consistent scalar density.  To obtain the standard RPA case, one 
simply has to replace  the self-consistent scalar density   
$\langle\phi^2\rangle_R$ by the Gaussian one   $\langle\phi^2\rangle_\varepsilon$ 
{\it i.e.} the expectation value is calculated on the vacuum of particles having 
the  energy $\varepsilon_{\vec q}$ and such that 
$\langle b^\dagger_{\vec q}\, b_{\vec q}\rangle_\varepsilon=0$. In one spatial 
dimension, the generalized mean-field mass $m$ is rendered finite by a simple mass 
renormalization:
\begin{equation}
m^2=\mu^2\,+\,{b\over 2}\,s^2\,+\,{b\over 2}\,\left(\langle\phi^2\rangle_R
\,-\,\int_{-\Lambda}^{+\Lambda}\,{dq\over 2\pi}\,
{1\over 2\sqrt{q^2+\mu^2}}\right) \label{MFMASS}
\end{equation}
where $\mu$ is the renormalized bare mass of the theory. 

\noindent
The r-RPA single particle propagator has been derived in I and is given by:
\begin{equation}
G(E, {\vec P}) = \left(\,E^2\,-\,\varepsilon^2_{\vec P}\,-\, \Sigma(E,{\vec P})
\right)^{-1}\quad\hbox{with}\quad\Sigma(E,{\vec P})={b^2\, s^2\over 2}\,
{I(E,{\vec P})\over 1\,-\,{b\over 2} I(E,{\vec P})}.
\end{equation}
The two-particle loop $I(E,{\vec P})$ 
has the explicit form given by  eq. (\ref{TWOLOOP}) in the appendix. It
explicitly depends on the ``occupation number''  ${\cal N}_{\vec
q}=\langle\phi_{\vec q} \, \phi^\dagger_{\vec q}\rangle_R$ which constitutes the
remaining problem to solve. One serious difficulty is that covariance is lost 
in the r-RPA in the sense that the loop integral 
$I(E, {\vec P})$ and consequently the mass operator $\Sigma(E, {\vec P})$
depends separately on $E$ and ${\vec P}$ due to the presence of the density 
${\cal N}$ in its expression. This is certainly a weakness of the present approach (see
discussion in I). One natural possibility  to recover covariance consists in 
imposing that the correct $I(E, {\vec P})$ is obtained through its center of mass (CM) 
expression according to: 
\begin{equation}  
I(E, {\vec P}) \equiv  I(E^2\,-\,{\vec P}^2)
= \int \,{d {\vec t}\over( 2\pi)^d}\,
{2{\cal N}_{t}\over E^2\,-\,{\vec P}^2\,-\,4\,\varepsilon_{t}^2\,+\, i\eta}~.
\end{equation}
The densities ${\cal N}_{t}$ can be calculated self-consistently
using the spectral theorem:
\begin{equation}
{\cal N}_{\vec P} =\int {i dE\over 2\pi}\,e^{i E \eta^+}\, G(E, {\vec P}).
\label{SPEC}
\end{equation} 
In the quasi-particle approximation used in I, the solution is  
${\cal N}_P=1/2\Omega_P$ where $\Omega_P=\sqrt{M^2 \,+\,P^2}$ is the energy of 
the one-particle RPA mode. As explained in details in I, the problem reduces to 
determine the pole of the one-particle propagator. One  important result of I is linked 
to the fact that the instability present in
$1+1$ dimensions at the level of the standard RPA (imaginary solution for the mass $M$ of 
the RPA mode) just disappears in r-RPA. 

\section{ r-RPA correlation energy in $\boldsymbol\Phi^4$ theory}
The energy of the system is calculated from the various Green's functions using the
spectral theorem. However, as is well-known, this cannot be done directly from the 
RPA results since important correlations would be lacking at the level of the expectation 
value  of the kinetic energy. To solve this problem we have generalized the so called 
``charging formula'' method \cite{FW} to the r-RPA case. We decompose the Hamiltonian in two 
pieces $\,H_0$ and $H_{int}$:
\begin{equation}
H =  V\left({1\over 2} \mu^2_0\,s^2 + {b\over 24}\,s^4\right)\,+\,
\sum_1\,{1\over 2}\left(\Pi_1\,\Pi^\dagger_1 + {\cal O}^2_1\,
\phi_1\phi^\dagger_1\right)\,+\,H_3\,+\,H_4 \equiv H_0 \,+\,H_{int} \label{HAM}
\end{equation}
where $H_3$ and $H_4$ are the 3- and 4-particle pieces of the Hamiltonian (last line 
of eq.~(\ref{HAM0})). $H_0$ has a form of a free Hamiltonian for quasi-particles with 
the generalized mean-field mass $m$ (eq.~(\ref{MFMASS})):
\begin{equation}
H_0=E_0\,+\,\sum_1\,{1\over 2}\left(:\Pi_1\,\Pi_1^\dagger:_\varepsilon\,+\,
\varepsilon^2_1\,:\phi_1\,\phi^\dagger_1:_\varepsilon\right)~.\label{H0}
\end{equation}
where $E_0$ is the generalized mean-field vacuum energy given in eq.~(61) of I.
As explained in subsection 4.2 of I, the interacting Hamiltonian is thus: 
\begin{equation}
H_{int}=H_3\,+\,H_4\,-\,{b\over 4}\,\langle\phi^2\rangle_R\,\sum_1\,
:\phi_1\,\phi^\dagger_1:_\varepsilon\,-\,V\,{b\over 8}\,
\langle\phi^2\rangle_\varepsilon^2~.\label{HINT}
\end{equation}
We also introduce an auxiliary Hamiltonian:
\begin{equation}
H'(\rho)\,=\,H_0\,+\,\rho\,H_{int}\, ,\qquad H'(\rho=1)=H~.
\end{equation}
The price to pay is  to solve the r-RPA problem (in practice  the
calculation of the commutators and 
double-commutators entering the RPA equations) for the $H'(\rho)$ Hamiltonian.
As explained in I, this can be done by making the following modifications
in the corresponding r-RPA problem for $H$:
\begin{eqnarray}
H\,&\to & \,H'(\rho)\nonumber\\
\varepsilon^2_1\,\to \,\varepsilon^2_{1\rho} &=&
\varepsilon^2_1\,+\,{b\over 2}\,\rho\,\left(\langle\phi^2\rangle_{R\rho}\,-\,
\langle\phi^2\rangle_R\right)\nonumber\\
H_3+H_4\,&\to &\,\rho\,(H_3\,+\,H_4) \label{MOD}
\end{eqnarray}
where $\langle\phi^2\rangle_{R\rho}$ is the self-consistent 
scalar density in the correlated RPA ground state of $H'(\rho)$.
We now employ the charging formula to calculate the ground-state energy as a
function of the condensate $s$, {\it i.e.,} the effective potential needed to
study the phase structure of the theory:
\begin{equation}
E_{RPA}=E_0\,+\,\int_0^1\, {d\rho\over\rho}\, \langle \rho\,H_{int}\rangle_\rho.
\label{ETOT}\end{equation}
Using Wick theorem with respect  to the vacuum of  the quasi-particle with energies 
$\varepsilon_\rho$, the correlated part can be rewritten as:
\begin{eqnarray}
\langle \rho\,H_{int}\rangle_\rho &=& 
\langle \rho \,H_3\rangle_\rho\,+\, \langle \rho\,
:H_4:_{\varepsilon_\rho}\rangle_\rho
-\,V\,{\rho\,b\over 8}\,\left(\langle \phi^2\rangle_{\varepsilon_\rho}\,-\,
\langle \phi^2\rangle_\varepsilon\right)^2\nonumber\\
& &-\,V\,{\rho\,b\over 4}\,\left(\langle \phi^2\rangle_R\,-\,
\langle \phi^2\rangle_{\varepsilon_\rho}\right)
\left(\langle \phi^2\rangle_{R\rho}\,-\,
\langle \phi^2\rangle_\varepsilon\right)~.\label{HINT2}
\end{eqnarray}
In this formula $\langle\phi^2\rangle_R$ is as before  the self-consistent scalar 
density of the original $H$  whereas $\langle \phi^2\rangle_{R\rho}$ corresponds 
to the same  quantity in the $H'(\rho)$ problem. $\langle\phi^2\rangle_{\varepsilon}$ 
is the scalar  density in the generalized mean field vacuum (vacuum of quasi-particles 
with energy $\varepsilon_{\vec q}$) in the $H$ problem and 
$\langle\phi^2\rangle_{\varepsilon_\rho}$ corresponds to the equivalent
quantity for the $H'(\rho)$ Hamiltonian. The  expectation values
$\langle\,\rho\,H_3\,\rangle_\rho$ and $\langle\,\rho\,:H_4:\,\rangle_\rho$ are 
calculated  by using the spectral theorem applied to the 1-2 and 2-2 Green's
functions relative to the r-RPA $H'(\rho)$ problem. These are actually the 
main contributions noted ${E^{(3)}}_{corr}$ and ${E^{(4)}}_{corr}$ of the correlation 
energy. However, at this level, 
we would like to precise one point. In the previous article I, we have taken into account
the term:  \label{FT}
\begin{eqnarray}
FT &=& \int\, \frac{d\rho}{\rho}\, 
\bigg[-\,V\,{\rho\,b\over 8}\,\left(\langle \phi^2\rangle_{\varepsilon_\rho}\,-\,
\langle \phi^2\rangle_\varepsilon\right)^2\nonumber \\
&&-\,V\,{\rho\,b\over 4}\,\left(\langle \phi^2\rangle_R\,-\,
\langle \phi^2\rangle_{\varepsilon_\rho}\right)
\left(\langle \phi^2\rangle_{R\rho}\,-\,
\langle \phi^2\rangle_\varepsilon\right) \bigg]
\end{eqnarray}
(called factorized term thereafter) in the following manner.
The self-consistent scalar density in the $\rho$ problem was calculated
by using the spectral theorem: 
\begin{eqnarray}
\exval{\phi^2}_{R \rho} = 
\int \frac{dp}{2\pi}\,\int \,\frac{i\,dE}{2\pi}\, e^{i E \eta^+}\, G_\rho(E,p).
\end{eqnarray} 
In this article, we calculate the difference of scalar densities for the various masses 
with the following formula ($\Lambda \rightarrow +\infty $):
\begin{eqnarray*}
\exval{\Delta\phi^2}_{m_1,m_2} &=& \exval{\phi^2}_{m_2} - \exval{\phi^2}_{m_1} 
          = \int_{-\Lambda}^\Lambda\ \frac{dp}{2\pi} \left( \frac{1}{2\sqrt{m_2^2 + p^2}}
          - \frac{1}{2\sqrt{m_1^2 + p^2}} \right) \\
          &=&  -\frac{1}{4\pi} \ln \left(\frac{m_2^2}{m_1^2}\right).
\end{eqnarray*}
We will show in the section about numerical results that this method gives better results.
In particular, with the non covariant terms (see discussion below)
we obtain this very important new result : the Simmon-Griffith theorem \cite{preuveMath} 
(which states that the order of the transition in the $\Phi^4$ model cannot be of 
the first order) is satisfied.

\bigskip\noindent
Two problems arise in the calculation of the effective potential.
First, there is an ambiguity in the calculation of the expectation
value $\langle\,\rho\,H_3\,\rangle_\rho$ which was not addressed in I and second,
non covariant terms appear, even when the covariance is forced in the
loop integral $I_\rho (E,P)$. We will show below  that the first
ambiguity gives a numerically   negligible effect and concentrate mainly on the 
non covariance problem. We notice that we have combinations of 
Green's function (see eqs.~(\ref{combi covariante G3}) 
and (\ref{combi covariante G4}) in the appendix) which are explicitly covariant
if the two-particle loop integral is covariant.
Hence, we can decompose the 3-particle and 4-particle correlation energies in
covariant pieces (which correspond to the expression (\ref{combi covariante G3}) and 
(\ref{combi covariante G4})) 
and non covariant pieces (the remaining terms) according to:
\begin{equation}
E_{corr}=E^{(3c)}_{corr}\,+\,E^{(3nc)}_{corr}\,+\,E^{(4c)}_{corr}\,+\,E^{(4nc)}_{corr}.
\end{equation}
The 4-body pieces have been given in I with the result~:
\begin{equation}
{E^{(4c)}_{corr}\over V}= \int_0^1{d\rho\over\rho}\,\int {d{\vec P}\over (2\pi)^d }\,
\int\,{i \,dE\over (2\pi)}\,e^{i\,E\,\eta^+} \,
I^2_\rho(E\, ,\,\vec P)\,F_\rho(E\, ,\,\vec P) , \label{E4C}
\end{equation}
\begin{equation}
{E^{(4nc)}_{corr}\over V}= \int_0^1{d\rho\over\rho}\,\int {d{\vec P}\over (2\pi)^d }\,
\int\,{i \,dE\over (2\pi)}\,e^{i\,E\,\eta^+} \,4\, 
I^{(1)\,2}_\rho(E\, ,\,\vec P )\,F_\rho(E\, ,\,\vec P)\label{E4NC}
 \end{equation}
\begin{equation}
\hbox{with}\qquad F_\rho(E\, ,\,\vec P)={1\over 24}\left({\rho^2\,b^2\over
1\,-\,{\rho\,b\over 2}\,I_\rho(E,{\vec P})}\,+\,{\rho^2\,b^3\,s^3\,G_\rho(E, {\vec P})
\over \left(1\,-\,{\rho\,b\over 2}\,I_\rho(E,{\vec P})\right)^2}\right)
\end{equation}
where the indices $\rho$ mean that the quantities are related to the $H'(\rho)$ problem.
The non covariance comes from the presence of the  loop integral:
\begin{equation}
I^{(1)}(E\, ,\,\vec P )\int \,{d{\vec k}_1\,d{\vec k}_2\over (2\pi)^d}\,{\delta^{(d)}
\left({\vec P}-{\vec k}_1 - {\vec k}_2\right)\,\over 2\,\varepsilon_1\,\varepsilon_2}\,
{\varepsilon_1\,{\cal N}_1\,+\,\varepsilon_2\,{\cal N}_2\over 
E\,-\,\varepsilon_1\,-\,\varepsilon_2\,+i\, \eta}
\end{equation}
which is not covariant ({\it i.e.,} it depends separately on $E$ and ${\vec P}$) even if
covariance is forced by taking as before the CM expression. We have neglected 
$E^{(4nc)}_{corr}$ (which vanishes to leading order in the interaction) in our previous
numerical estimate in I. However this contribution, although relatively small, turns out to
be very important for the precise nature of the phase transition.  In other words 
$E^{(4nc)}$ is very important to reproduce good numerical results.  Lets us come to the
calculation of the expectation value of $H_3$:
\begin{eqnarray}
\langle H_3\rangle &= &{b\,s\over 6\, \sqrt{V\,\Pi_i 2\varepsilon_i}}
\,\delta_{1+2+3}\,\bigg(
\langle (b^\dagger_1\, b^\dagger_2\,+\, b_{-1}\, b_{-2})\,
(b_{-3}\,+\, b^\dagger_3)\nonumber\\
& & +\,2\,(b^\dagger_1\, b^\dagger_2\, b_{-3}
\,+\,b^\dagger_1\,b_{-2}\, b_{-3})\rangle\bigg)\label{AMB}.
\end{eqnarray}
The first line generates the covariant contribution: 
\begin{equation}
{E^{(3c)}_{corr}\over V}=\int_0^1{d\rho\over\rho}\,\int {d{\vec P}\over (2\pi)^d }\,
\int\,{i \,dE\over (2\pi)}\,e^{i\,E\,\eta^+}\,
{\rho^2\,b^2\,s^2\over 6} \,
{G_\rho(E,\vec P)\,I_\rho(E,{\vec P})\over 1\,-\,{\rho\,b\over 2}\,I_\rho(E,{\vec P})}
\label{E3C}
\end{equation}
which was already considered in I. The second line of eq. (\ref{AMB}) 
generates the already
mentioned ambiguity: it is not uniquely defined since two different combinations of
Green's functions can be used:
\begin{equation}
\delta_{1+2+3} \exval{\bd_1\ob_{-2}\ob_{-3}}  
  =\delta_{1+2+3} \,\int \frac{i\,dE}{2\pi} 
e^{\imath E\eta^+} \G^{(1)}_{-2-3, 1^\dagger}(E)
\label{non zero}
\end{equation}
or
\begin{equation}
 \delta_{1+2+3} \exval{\bd_1\ob_{-2}\ob_{-3}} 
   = \delta_{1+2-3}\,\int \frac{i\,dE}{2\pi} e^{\imath E\eta^+}
   \G^{(2)}_{-1-2, 3}(E) 
\label{zero}
\end{equation}
where the notations of the appendix (eq. (\ref{A12})) have been used. The second form 
gives identically zero in standard RPA and in r-RPA when covariance is forced in
$I^{(2)}$ (see appendix for its definition),
whereas the first one gives a finite contribution. If eq. (\ref{non zero}) is
used, we obtain for the non covariant correlation energy:
\begin{eqnarray}
{E^{(3nc)}_{corr}\over V}
  &=& \int_0^1{d\rho\over\rho}\,\int {d{\vec P}\over (2\pi)^d }\,
\int\,{i \,dE\over (2\pi)}\,e^{i\,E\,\eta^+} \, {\rho^2\,b^2\,s^2\over 6}\,
\left( \frac{E + \epsilon_{\rho P}}{2\epsilon_{\rho P}} \right)\, 
\frac{2\, I^{(1)}_\rho(E\, ,\,\vec P )\,G\rho(E,\vec P)}
{1 - \frac{\rho\, b}{2}I_\rho(E,\vec P)} . \nn \\
\label{E3NC}
\end{eqnarray}

\begin{figure}[ht]
\begin{center}
\begin{tabular}{cc}
\includegraphics[width=0.4\linewidth]{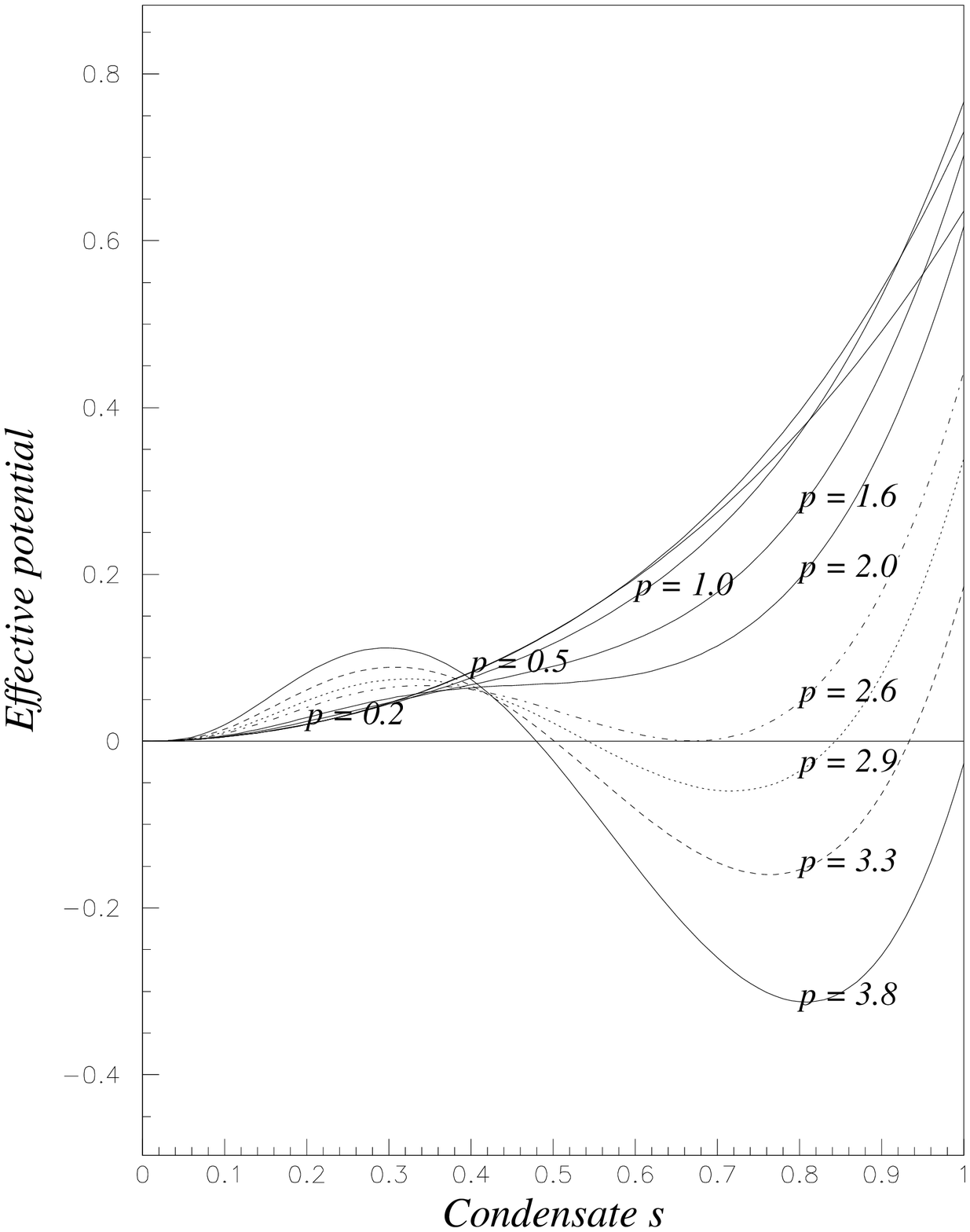} & 
\includegraphics[width=0.4\linewidth]{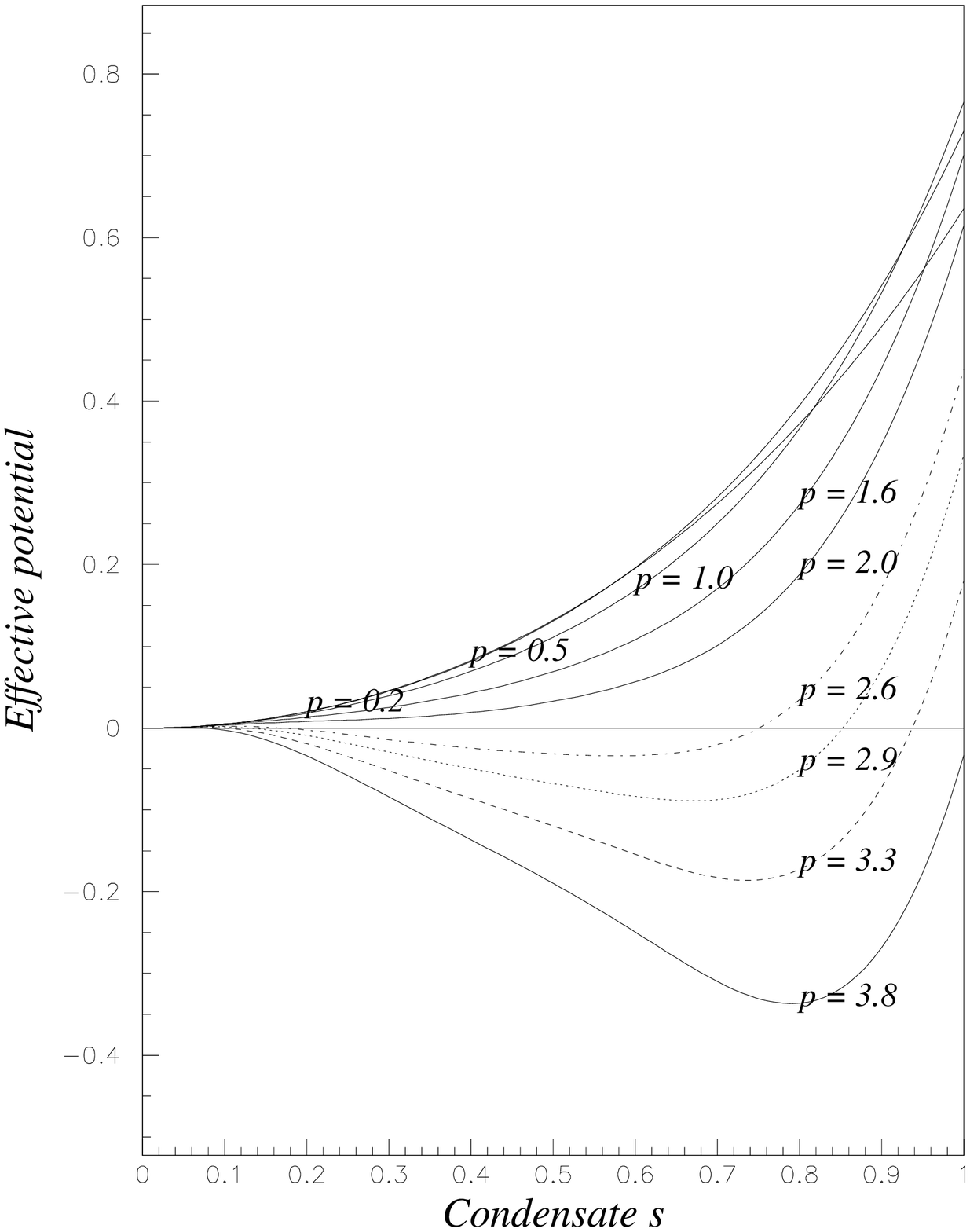}
\end{tabular}
\caption{Generalized mean field effective potential (left panel); 
generalized mean field effective potential plus the factorized term $FT$ (right panel),
for different values of the reduced coupling constant $p$, as a function of $s$.}
\label{fig:GMF+corr}
\end{center}
\end{figure} 

In $1+1$ dimensions all the various contributions can be calculated using a Wick rotation
($E^2\,-\,P^2\to-{\cal S}$) according to the method explained in I. 
\begin{eqnarray}
{E^{(4c)}_{corr}\over V} &=&-\int_0^1\, {d\rho\over\rho}\,
\int_0^\infty\,{d{\cal S}\over 4 \pi }\,
I^2_\rho(-{\cal S})\,F_\rho(-{\cal S})\\
{E^{(3c)}_{corr}\over V} &=&-\int_0^1\, {d\rho\over\rho}\,
\int_0^\infty\,{d{\cal S}\over 4 \pi }\,
{\rho^2\,b^2\,s^2\over 6} \,
{I_\rho(-{\cal S})\,G_\rho(-{\cal S})\over 1\,-\,{\rho\,b\over 2}
\,I_\rho(-{\cal S}))}
\end{eqnarray}
\begin{eqnarray}
{E^{(4nc)}_{corr}\over V} &=&\int_0^1{d\rho\over\rho}\,
\int_0^\infty {d{\cal S}\over 2\pi^2}\,\int_0^{\pi/2}d\theta
\bigg({\cal S}\,\cos^2\theta\,J^2({\cal S}\,,\,\theta)\,-\,I^2(-{\cal S})
\bigg)\,F_\rho(-{\cal S})\\
{E^{(3nc)}_{corr}\over V}
  &=& -\int_0^1\, {d\rho\over\rho}\, \frac{\rho^2b^2s^2}{3}\, 
  \int\,  {d{\cal S}\over 2\pi^2 }\, \nn \\
& &\times  \int_0^{\frac \pi 2}\, d\theta\, 
\left( 
\frac{ {\cal S}\cos^2\theta }{ 4 \sqrt{\varepsilon_{ t\rho}^2 + {\cal S}\sin^2\theta} } J({\cal S},\, \theta)
- \frac 1 4 I(-{\cal S})
\right)
\frac{G_\rho(-{\cal S})}{1 - \frac{\rho\, b}{2}I_\rho(-{\cal S})} 
\label{E3NC Wick}
\end{eqnarray}
with
\begin{equation}
\label{J la}
J({\cal S}\,,\,\theta)=-\int \,{d t\over 2\pi}\,
{1\over \sqrt{4 \varepsilon_{ t\rho}^2\,+{\cal S} \sin^2\theta}}\,
{2{\cal N}_{ t\rho}\over {\cal S}\,+\,4\,\varepsilon_{ t\rho}^2}~.
\end{equation}

\begin{figure}[ht]
\begin{center}
\begin{tabular}{cc}
\includegraphics[width=0.4\linewidth]{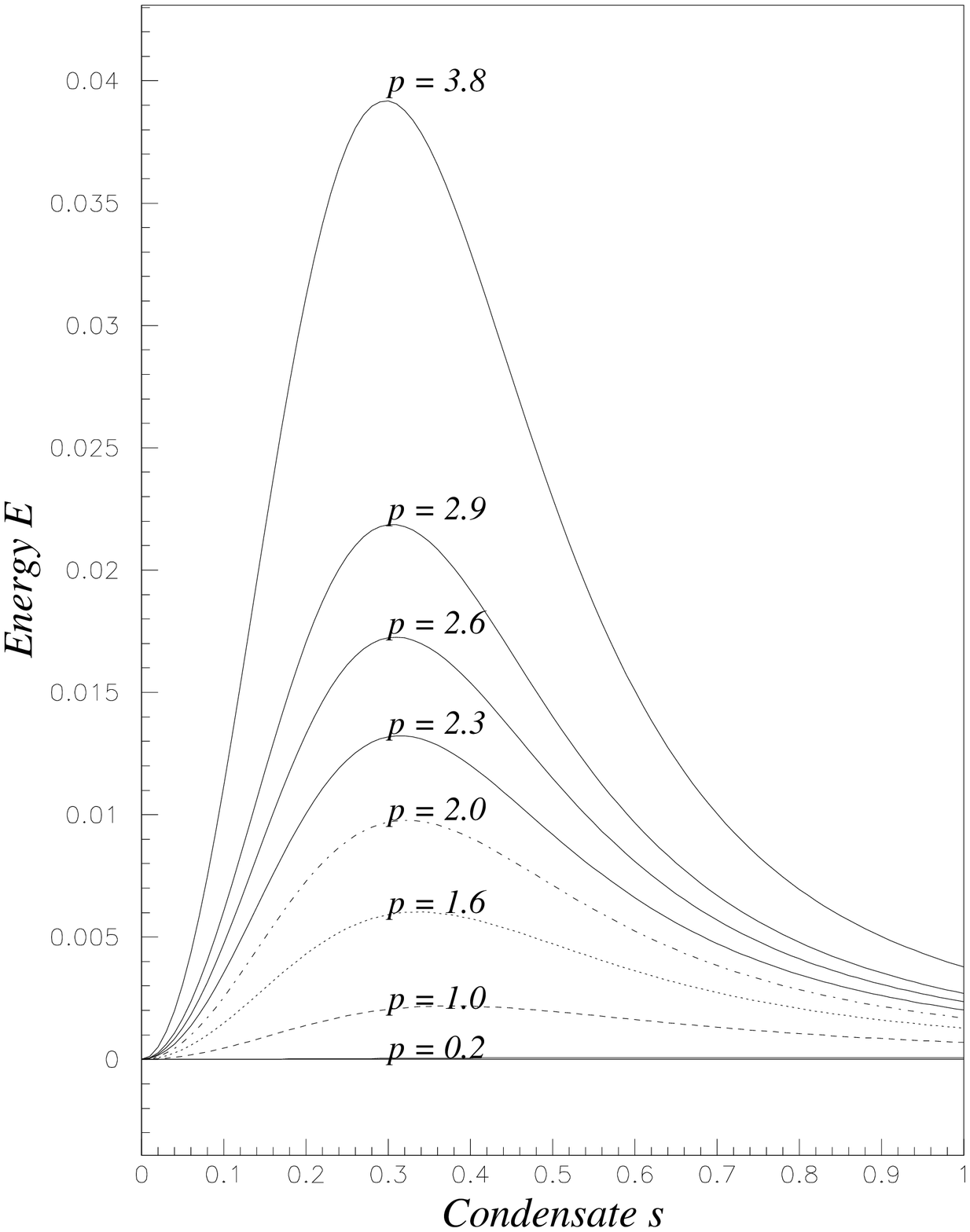} & 
\includegraphics[width=0.4\linewidth]{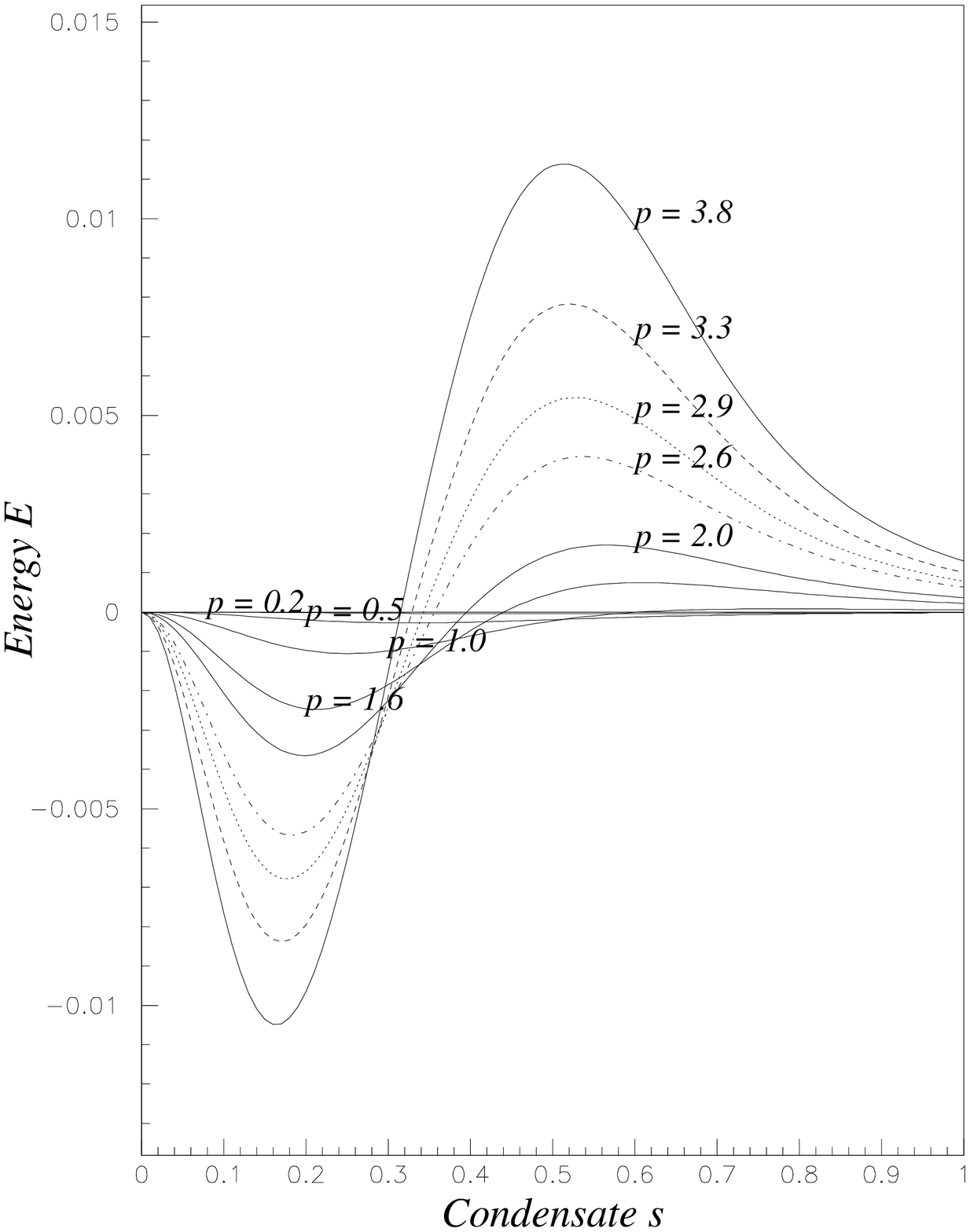}
\end{tabular}
\caption{Non covariant 3-body correlation energy $E^{(3nc)}$ calculated with eq. (\ref{E3NC Wick})
(left panel: standard RPA, right panel: r-RPA), for 
different values of the reduced coupling constant $p$, as a function of $s$.}
\label{fig:E3NC}
\end{center}
\end{figure} 

\section{Numerical results}
Before going to the discussion of the correlation energy, let us mention
the following result. 
The generalized mean-field effective potential $E_0(s)$ 
(vacuum energy as a function of
the condensate $s$) presents a strong first order phase transition
as shown on fig. \ref{fig:GMF+corr}, left panel (all numerical results
are obtained with the reduced coupling constant $p = b/24\mu^2$ and $\mu = 1$).
When we add the factorized term $FT$,
this strong potential barrier disappears (fig. \ref{fig:GMF+corr}, right panel).
The transition is practically, up to an extremely small potential barrier, of second
order nature with a critical coupling constant $p_c\simeq 2.3$. It is thus tempting to
consider the sum $E_0 + FT$ as  the true mean field energy for
the r-RPA calculation because it contains all factorisable or reducible parts of
the total energy.  The fact that this term has a second order phase transition is very 
important in the following. The full energy is obtained by adding the
pure interaction terms $\exval{H_3+:H_4:}$ (diagrammatically, irreducible terms) 
which we will discuss
below.

\begin{figure}[ht]
\begin{center}
\begin{tabular}{cc}
\includegraphics[width=0.4\linewidth]{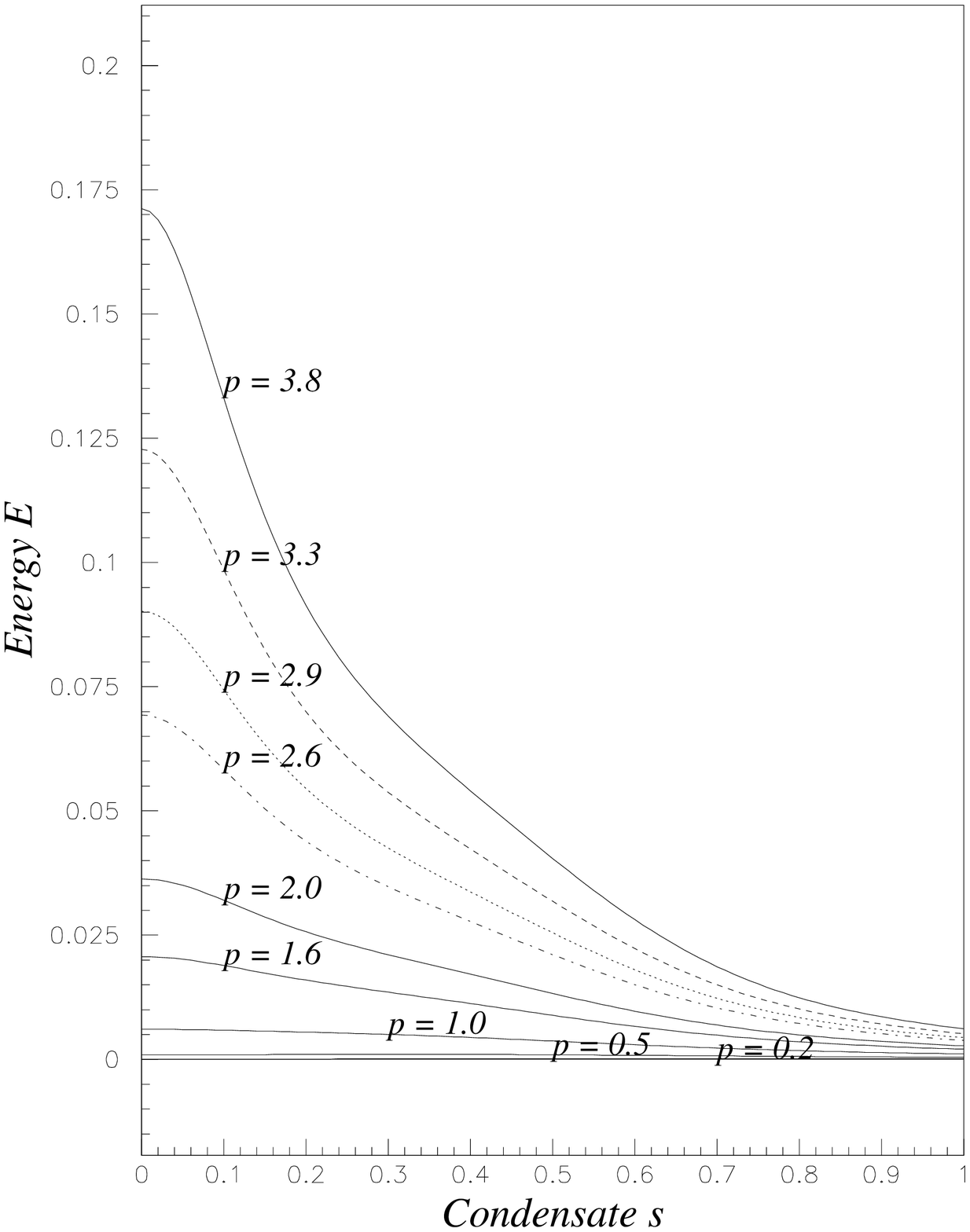} & \includegraphics[width=0.4\linewidth]{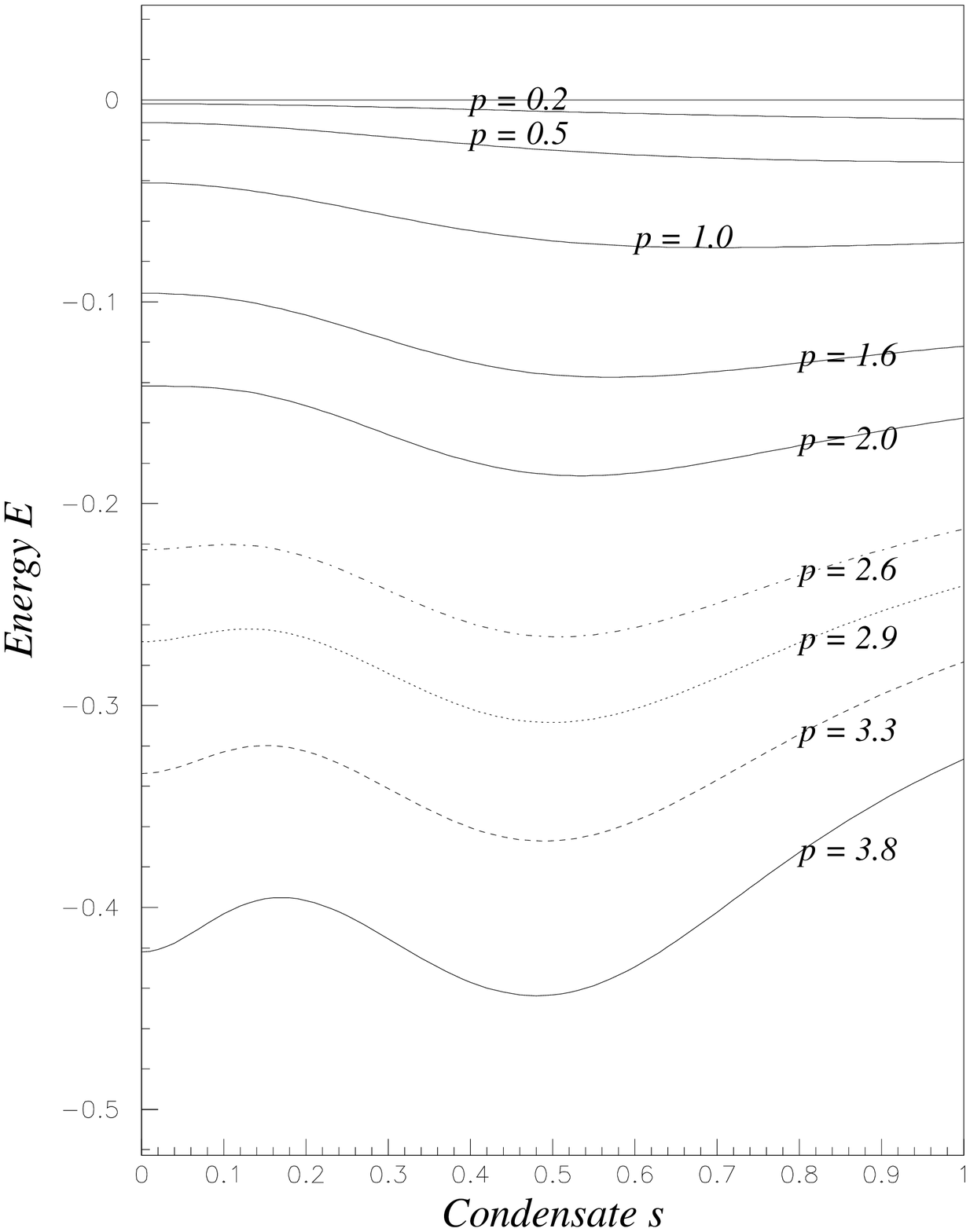}
\end{tabular}
\caption{Left panel: non covariant 4-particle correlation energy $E^{(4nc)}$. 
Right panel: covariant correlation energy $E^{(3c)} + E^{(4c)}$
(in r-RPA, for different values of $p$, as a function of $s$).}
\label{fig:Ecorr}
\end{center}
\end{figure} 

\begin{figure}[ht]
\begin{center}
\begin{tabular}{cc}
\includegraphics[width=0.4\linewidth]{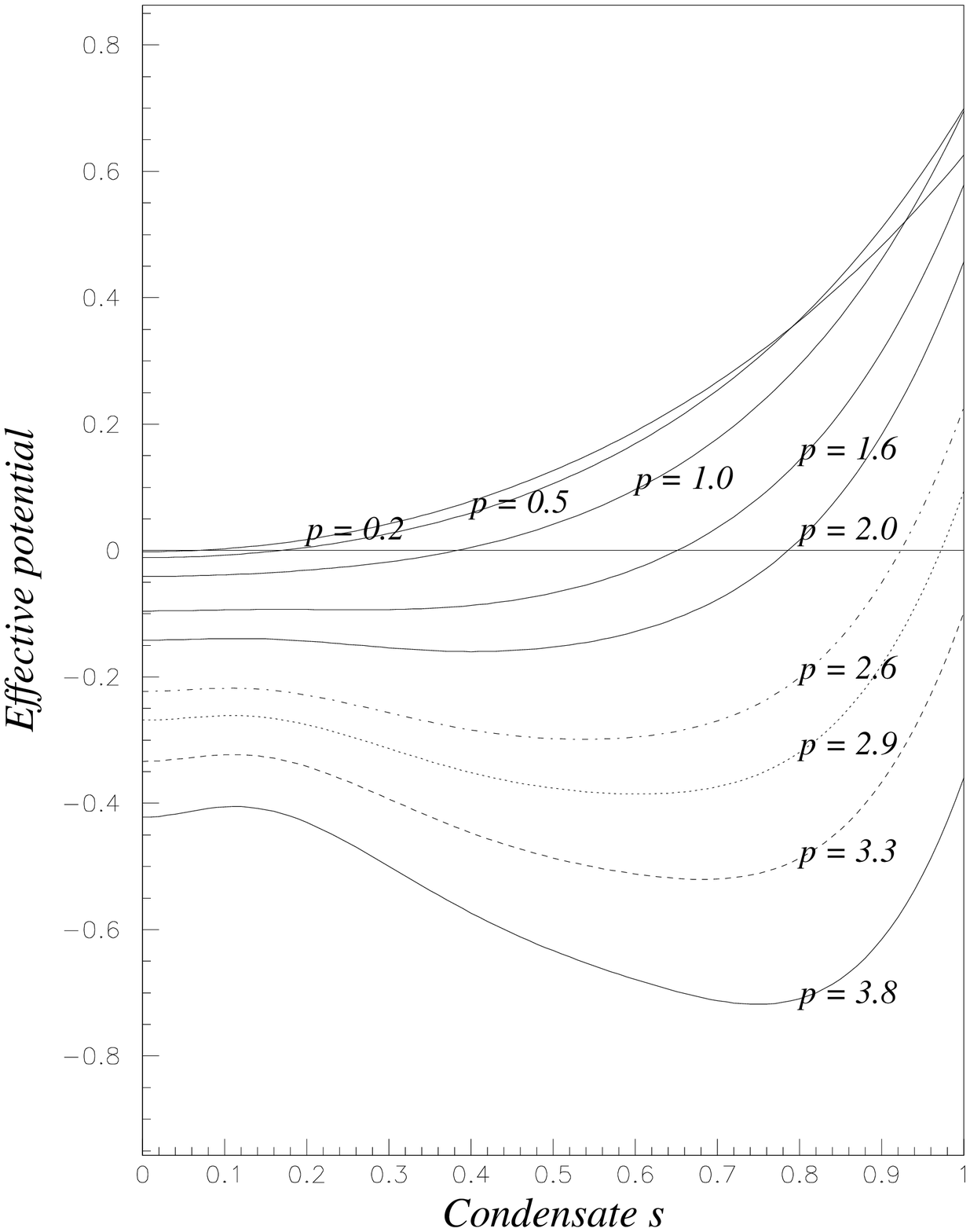} 
& \includegraphics[width=0.4\linewidth]{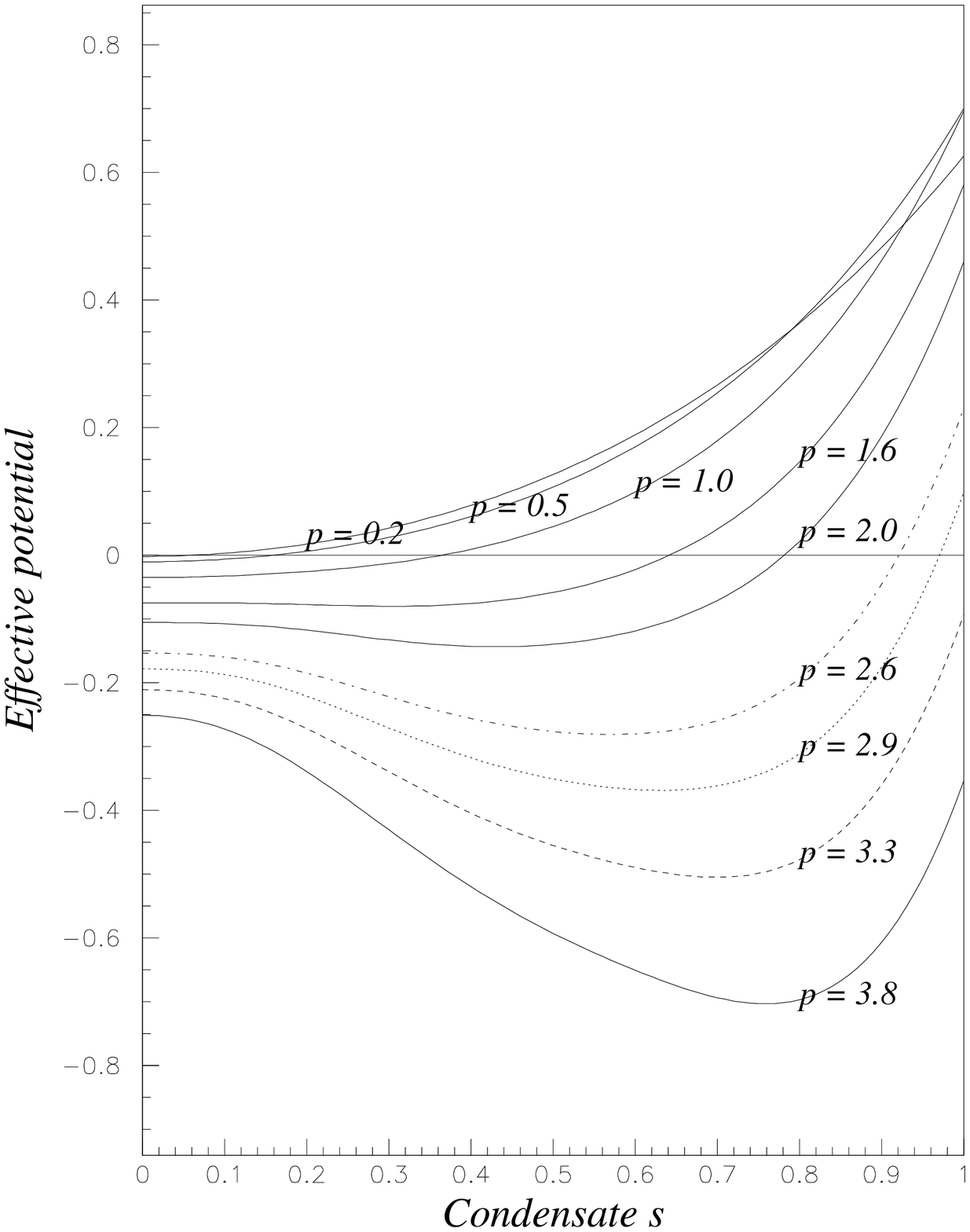}
\end{tabular}
\caption{Effective potential in r-RPA;  Left panel: covariant potential 
$E_{tot}^{(c)} = E_0 + E^{(3c)} + E^{(4c)} + FT$.
Right panel: total potential $E_{tot} = E_0 + E^{(3c)} + E^{(4c)} + E^{(4nc)} + FT $.}
\label{fig:E rRPA}
\end{center}
\end{figure} 

\begin{figure}[ht]
\begin{center}
\begin{tabular}{cc}
\includegraphics[width=0.4\linewidth]{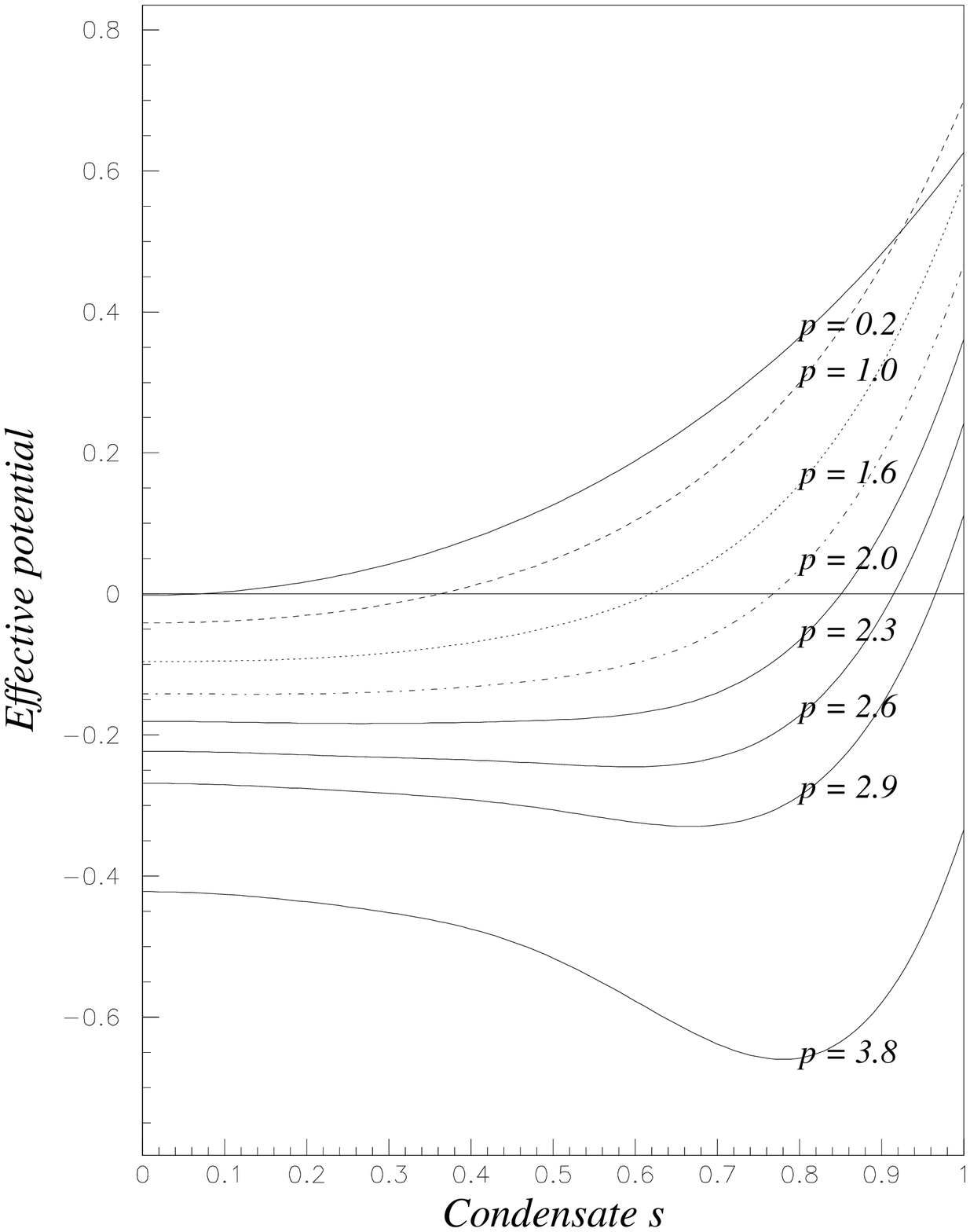} & 
\includegraphics[width=0.4\linewidth]{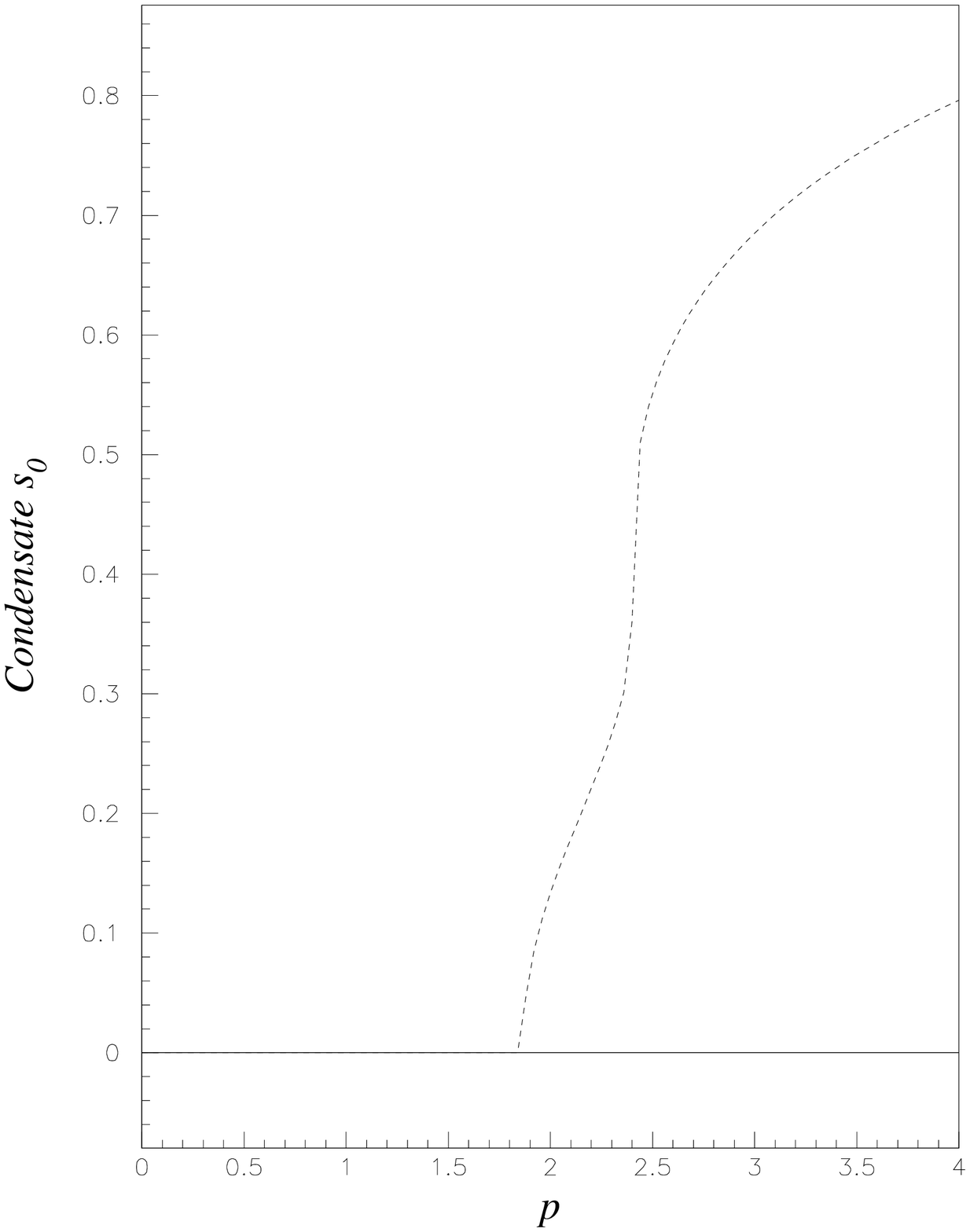}
\end{tabular}
\caption{Left panel: Covariant effective potential in s-RPA $E_{tot}^{(c)} = E_0 + E^{(3c)} 
+ E^{(4c)}$.
Right panel: value of the condensate at the global minimum  of $E_{tot}^{(c)}$ in s-RPA. 
The transition is of the second order,
because this curve is continuous.}
\label{fig:E sRPA}
\end{center}
\end{figure} 

We first consider the ambiguous term $E^{(3nc)}$ term using eq. (\ref{E3NC Wick})
and we obtain the results
shown on fig. \ref{fig:E3NC} for standard and renormalized RPA.
We can verify that this term is negligible at least in r-RPA. 
The maximum of this term is about respectively $10$ 
and $100$ times in s-RPA and r-RPA, smaller than the  covariant contribution  
in the correlation energy shown on fig. \ref{fig:Ecorr}. This demonstrates 
that the ambiguity linked to the non covariance problem is not so serious in r-RPA. 
Hence, in  the following we take the reasonable option of putting this contribution 
to zero. The other term not considered in I,  $E^{(4nc)}$, is shown on 
fig. \ref{fig:Ecorr} in comparison with the covariant correlation energy
$E^{(3c)} + E^{(4c)}$. We clearly see that $E^{(4nc)}$ which was ignored in I is 
sizeable especially for low values of the condensate $s$.
The covariant effective potential ($E_{cov} = E_0 + FT + E^{(3c)} + E^{(4c)} $)
and the total effective potential ($E_{tot} =  E_0 + FT + E^{(3c)} + E^{(4c)} + E^{(4nc)}$) 
in r-RPA 
are shown on  fig. \ref{fig:E rRPA}. For completeness we also show on fig. \ref{fig:E sRPA} 
the  results   obtained in standard RPA which were already given in I 
(in this calculation the RPA single particle propagator is
replaced by the mean field one to avoid the divergence associated with instability).
One sees that the RPA fluctuations in the standard RPA case are 
able to transform the strong first order transition of the 
Gaussian approximation into a second order phase transition.  
This comes from the fact that the attractive s-RPA correlations in the metastable
region of the Gaussian approximation ($p\in [0.2, 0.5]$) strongly reduce  the potential
barrier of the Gaussian effective potential.
We see on fig. \ref{fig:E sRPA}
the  evolution of the global minimum of the effective potential  with increasing $p$. The fact 
that the transition is of second order is demonstrated by the fact that the condensate $s$ is 
a continuous function of $p$ and this is in agreement with the Simon-Griffiths theorem 
\cite{preuveMath} which states that
the phase transition in the $\lambda \phi^4$ cannot be of the first
order. However the critical coupling $p_c=1.8$
is different from both the lattice result \cite{JS94,LW97,MRL99},  
$p_c=2.55$, and the cluster expansion technique result \cite{HCPT95}, $p_c=2.45$. 
Although this result is not so bad, the
standard RPA result cannot be really trusted since this method is spoiled by the instability
problem. We obtained in I a preliminary result in r-RPA but keeping only the covariant pieces in
the correlation energy. Here, the result is shown on the left panel of fig. 
\ref{fig:E rRPA} with the explicit incorporation  of the FT term as explained in 
 section \ref{FT}. 
Although much less marked than in the Gaussian case, one again obtains a first
order transition. There is nevertheless a slight progress with respect  to the result obtained in I. 
The potential barrier is smaller and
the critical coupling constant,
$p_c=1.9$, is much closer to the lattice and cluster results. When the non covariant contribution 
$E^{(4nc)}$ is added, the transition is of  second order nature (see fig. \ref{fig:E rRPA}, right panel).
Indeed, this repulsive $E^{(4nc)}$ decreases 
the correlation energy in the meta-stable region {\it i.e.,} around the local maximum. 
Consequently the potential barrier is so 
weakened that it becomes negligible (about $10^{-3}$). Although the critical parameter is $p_c = 1.6$,
this constitutes an important result of this paper. 
The restoration of Simmon-Griffith theorem indicates that r-RPA describes
correctly the phase transition region. In addition the absolute value
of the effective potential becomes very similar to the cluster effective potential.  
As an example  for $s=0$ and $p=2$, one gets the following results for the total energy. 
Ignoring the non covariant contribution one gets $-0.14$ to be compared with $-0.11$ in the
cluster calculation. When  $E^{(4nc)}$ is added, one gets $-0.105$. This is a very
encouraging result and certainly further work is needed in direction of the fully self-consistent
RPA. The problem of the covariance is certainly a key issue in that respect. One interesting
possibility is to incorporate three-body excitations in the line of the work of ref.
\cite{DHO02}. 

\section{The linear$-\sigma$ model}
We now discuss another model, the \linsig\ one.
It can be seen as a Ginsburg-Landau effective Lagrangian for the $SU(2)\times SU(2)$ 
chiral symmetry (spontaneously broken in one direction).
This model can give us physical insight in the chiral phase transition and may also describe
far from the transition the dynamics of pionic systems.\\
The Lagrangian reads:
\begin{equation}
 {\cal L}_{\sigma} =  \frac 1 2 \partial_\mu\sigma\partial^\mu\sigma + 
\frac 1 2 \partial_\mu\vec\pi\partial^\mu\vec\pi 
 - \frac{m^2} 2 \left( \sigma^2 + \pi^2 \right) - \frac \lambda 4 {\left( \sigma^2 + \pi^2 \right)}^2 + c\sigma
\label{lagrangien siglin}
\end{equation}
where $\vec\pi$ is pseudo-scalar isovector field and $\sigma$ a scalar iso-scalar field.
$\vec\pi$ corresponds to the physical pion and its  chiral partner $\sigma$ may describe 
a mode associated
with the amplitude fluctuation of the chiral condensate.

\noindent
This model can formally be seen as a generalization of the $\Phi^4$ model for a \mbox{$N+1$} dimensional
multiplet $(\sigma,\vec\pi)$. It possesses an exact  $O(N+1)$ invariance  if the parameter $c$ is zero.
The  $c\sigma$ piece of the Lagrangian describes the amount of  explicit breaking  of chiral  
symmetry in QCD.

\noindent
For the application of r-RPA we use notations similar to the $\Phi^4$ model.
We introduce $s$, the condensate in  $\sigma$ field direction (the chiral symmetry being broken in
one direction) and the fluctuating field $\sigma'$: 
$\sigma = \sigma' + s$ with $\exval\sigma = s$ (hence $\exval{\sigma'} = 0$) and we
omit the prime thereafter. We note $\Pi_\sigma$ and $\vec\Pi_\pi$ the conjugate momenta of
the fields $\sigma$ and $\vec\pi$ with usual commutation relations.

\noindent
The Hamiltonian reads: 
\begin{align}
  H(\sigma,\pi) &= \int\, dx\, 
\bigg\{
\frac 1 2 \big( \vec\Pi^2_\pi + (\vec\nabla\vec\pi)^2 + (\mu_0^2 + \lambda s^2)\vec\pi^2 \big) 
+ \frac 1 2 \big( \Pi^2_\sigma + (\nabla\sigma)^2 + (\mu_0^2 + 3 \lambda s^2)\sigma^2 \big) \nn \\
& + \lambda s \sigma (\sigma^2 + \vec\pi^2) 
+ \frac \lambda 4 { (\sigma^2 + \vec\pi^2)}^2
+ \sigma (\mu_0^2 s + \lambda s^3 - c)
+ \frac{\mu_0^2} 2 s^2 + \frac \lambda 4 s^4 - c s
\bigg\} .
\end{align}
We define quasi-particle operators for quasi-pion and quasi-sigma:
$$  {\op{b}_\pi}_\beta
=\sqrt{{\kappa_\pi}_\beta\over 2}\,\pi_\beta\,+\,
  \sqrt{1\over 2 {\kappa_\pi}_\beta} {\Pi_\pi}_\beta\quad 
\mbox{ and }
  {\quad\op{b}_\sigma}_\beta
=\sqrt{{\kappa_\sigma}_\beta\over 2}\,\sigma_\beta\,+\,
  \sqrt{1\over 2 {\kappa_\sigma}_\beta} {\Pi_\sigma}_\beta
$$
and we introduce the scalar densities
${\cal N}_\alpha = \langle\pi_\alpha \pi_\alpha^\dagger\rangle$ and
$N_\alpha = \langle\sigma_\alpha \sigma_\alpha^\dagger\rangle$. For what concerns the sigma operators, the
indice $\alpha$ represents a momentum state whereas for the pionic operators it represents
a momentum and isospin state.\\

\noindent
As in ref. \cite{ACSW} we can introduce  the r-RPA excitation operators in the pionic channel
according to:
$$
 \op Q^\dag_\nu = 
      X_\alpha\ \op{b}^\dag_{\pi,\, \alpha} - Y_{-\alpha}\ \op{b}_{\pi, \, -\alpha}
      + X_{\alpha \beta}\ \op{b}^\dag_{\pi,\, \alpha} \op{b}^\dag_{\sigma,\, \beta} 
      -  Y_{-\alpha-\beta}\   \op{b}_{\pi,\, -\alpha} \op{b}_{\sigma,\, -\beta} 
$$
This is equivalent, in the Green's function approach that we really use at variance with
\cite{ACSW}, to  calculate the RPA correction to 
the pion mass operator ($\Sigma_\pi$) originating from the $\pi\sigma$ RPA bubbles.

\noindent
With the same techniques used in the $\Phi^4$ theory, we derive the generalized mean field equations
for pion and sigma modes. For $N=3$ they read:
\begin{align}
 \epsilon_\alpha^2 &= \mu_0^2 + k^2_\alpha +   \lambda \, \sum_\alpha (N_\alpha + 5 {\cal N}_\alpha) + \lambda s^2\\
 E_\alpha^2        &= \mu_0^2 + k^2_\alpha + 3 \lambda \, \sum_\alpha (N_\alpha +   {\cal N}_\alpha) + 3\lambda s^2 .
\end{align}
The r-RPA inverse pion propagator is obtained as:
\begin{equation}
 G^{-1}(\omega, k_\alpha) = \omega^2 - \epsilon^2_\alpha(\omega) - \Sigma^\pi_\alpha(\omega) .
\label{pion propagator}
\end{equation}
with the pion mass operator given by:
$$
 \Sigma^\pi_\alpha(\omega) = 4 \lambda^2 s^2 \frac{ I^{\pi\sigma}_\alpha(\omega) }
 {1-2\lambda I^{\pi\sigma}_\alpha(\omega) }.
$$
It contains an iteration of the $\pi\sigma$ bubble $I^{\pi\sigma}$ which is given by: 
\begin{equation}
  I^{\pi\sigma}_\alpha(\omega) = \sum_{\beta\beta'} 2\delta_{\alpha-\beta-\beta'} 
\frac{( {\cal N}_\beta + N_{\beta'}  )\omega^2 + ( N_{\beta'} - {\cal N}_\beta )(E^2_{\beta'} - \epsilon^2_\beta) }
{ [\omega^2 - (\epsilon_\beta + E_{\beta'})^2 ] [\omega^2 - (\epsilon_\beta - E_{\beta'})^2 ]  }
\end{equation}
or, in a form analogous to the RPA loop in $\Phi^4$ model~:
\begin{equation}
  I^{\pi\sigma}_\alpha(\omega) = \sum_{\beta\beta'}  \delta_{\alpha-\beta-\beta'} 
\frac{\epsilon_\beta + E_{\beta'}}{2\epsilon_\beta E_{\beta'}} \,
\frac{{\cal N}_\beta \epsilon_\beta + N_{\beta'} E_{\beta'}}
{\omega^2 - (\epsilon_\beta + E_{\beta'})^2}
- \frac{\epsilon_\beta - E_{\beta'}}{2\epsilon_\beta E_{\beta'}} \,
\frac{{\cal N}_\beta \epsilon_\beta - N_{\beta'} E_{\beta'}}
{\omega^2 - (\epsilon_\beta - E_{\beta'})^2} .
\end{equation}

\noindent
Our preliminary goal was to show that the r-RPA fluctuations restore the Goldstone theorem.
With the use of the generalized mean field equations and using the expression of 
eq. (\ref{pion propagator})
for $\omega = 0$, one can obtain the result:
\begin{align}
 G^{-1}(\omega, k_\alpha) &= \omega^2 - [\mu_0^2 + \lambda s^2 
+ 3\lambda \, \sum_\alpha (N_\alpha + {\cal N}_\alpha) ] 
- \big( \Sigma^{\pi}_\alpha(\omega) - \Sigma^{\pi}_0(\omega = 0) \big) \label{Geq1} \\
&= \omega^2 - \frac c s
- \big( \Sigma^{\pi}_\alpha(\omega) - \Sigma^{\pi}_0(\omega = 0) \big) \label{Geq2}
\end{align}
(the last result (\ref{Geq2}) is obtained by using the gap equation $\partial E/\partial s = 0$ 
which shows
that the term in bracket in eq.(\ref{Geq1}) is just $c/s$).
These expressions clearly show  that the Goldstone theorem is satisfied in the chiral limit because
the spurious mode $\omega = 0$ is allowed. \\

\noindent
As a preliminary conclusion, we underline this encouraging result: the r-RPA fluctuations can 
 correctly treat the 
spontaneously broken symmetry  and the spurious mode is obtained even if the covariance is lost. 
In particular it restores 
the Goldstone theorem violated at the level of the mean field or Gaussian approximation \cite{DMS96}. 
This generalizes in r-RPA the result  already obtained in the s-RPA formalism in \cite{ACSW}.

\section{Conclusion}
We have discussed  in this article some problems encountered in the calculation of
the effective potential in r-RPA and we have also presented some new numerical results for the
$\lambda \Phi^4$ theory in 1+1 dimensions. We have shown  that the  ambiguity in 
the calculation of the 3-particle  energy is only apparent in the sense that it is numerically
very small. We have also shown that the incorporation of the so-called non covariant 
contributions in the effective potential significantly improves the description of the 
phase transition in the direction of lattice and cluster expansion results. 
The most important result of this work is that we found a way to take into account the
different contributions of the RPA correlations that give us a second order phase 
transition.

For what concerns the \linsig\ model we have shown that the Goldstone theorem is explicitly satisfied
despite of the covariance problem. A further work of interest is evidently to calculate the effective
potential possibly at finite temperature to study the chiral phase transition.

\section{Acknowledgments}
We thank P. Schuck, D. Davesne, M. Oertel and A. Rabhi for constant interest in this work and many
fruitful discussions.
One of the author (H.Hansen) is supported by the grant SFRH/BPD/11579/2002 of FCT.

\section{Appendix}

In this appendix we list the explicit expressions for the  Green's functions.
Here the momentum indices are represented by greek letters $\alpha, \beta, \gamma$. 
For this purpose we introduce various quantities:
\begin{eqnarray}
I^{(1)}_{\beta\beta'}(E)& = &{ 1\over 2\varepsilon_\beta} {1\over 2\varepsilon_{\beta'}}
\,{\varepsilon_\beta\,{\cal N}_\beta\,+\,\varepsilon_{\beta'}\,{\cal N}_{\beta'}\over
E\,-\,\varepsilon_\beta \,-\,\varepsilon_{\beta'}\,+\,i\eta}\nonumber\\
I^{(2)}_{\beta\beta'}(E)& = &{\varepsilon_\beta\,-\,\varepsilon_{\beta'} 
\over 2\,\varepsilon_\beta \,\varepsilon_{\beta'}}\,
{\varepsilon_\beta\,{\cal N}_\beta\,-\,\varepsilon_{\beta'}\,{\cal N}_{\beta'}\over
E^2\,-\,(\varepsilon_\beta \,-\,\varepsilon_{\beta'})^2\,+\,i\eta}\nonumber\\
I^{(3)}_{\beta\beta'}(E)& = &-{ 1\over 2\varepsilon_\beta} {1\over 2\varepsilon_{\beta'}}
\,{\varepsilon_\beta\,{\cal N}_\beta\,+\,\varepsilon_\beta'\,{\cal N}_{\beta'}\over
E\,+\,\varepsilon_\beta \,+\,\varepsilon_{\beta'}\,-\,i\eta}~.\nonumber\\
\end{eqnarray}
We also introduce the loop integrals~;
\begin{eqnarray}
I^{(i)}_\alpha (E) &=& {1\over V}\sum_{\beta{\beta'}}\,\delta_{\alpha-\beta-\beta'}\,
I^{(i)}_{\beta\beta'}(E)\nonumber\\
I_\alpha (E) &=& I^{(1)}_\alpha (E)\,+\,I^{(2)}_\alpha (E)\,+\,I^{(3)}_\alpha (E)~.
\end{eqnarray}
In particular for $\alpha$ corresponding to the momentum ${\vec P}$ one has the explicit
expression:
\begin{eqnarray}
I(E,{\vec P}) &\equiv & I_{\alpha={\vec P}}(E)\nonumber\\
& = &\int \,{d{\vec k}_1\,d{\vec k}_2\over (2\pi)^d}\,\delta^{(d)}\left({\vec P}-
{\vec k}_1 - {\vec k}_2\right)\,\bigg[
{\varepsilon_1\,+\,\varepsilon_2\over 2\,\varepsilon_1\,\varepsilon_2}\,
{\varepsilon_1\,{\cal N}_1\,+\,\varepsilon_2\,{\cal N}_2\over 
E^2\,-\,\left(\varepsilon_1\,+\,\varepsilon_2\right)^2+i \eta}\nonumber\\
& &-\,
{\varepsilon_1\,-\,\varepsilon_2\over 2\,\varepsilon_1\,\varepsilon_2}\,
{\varepsilon_1\,{\cal N}_1\,-\,\varepsilon_2\,{\cal N}_2\over 
E^2\,-\,\left(\varepsilon_1\,-\,\varepsilon_2\right)^2+i \eta}\bigg]\label{TWOLOOP}.
\end{eqnarray}
For the one-particle Green's functions one obtains:
\begin {eqnarray}
G_{\alpha{\alpha'}^\dagger}(E) &=&\delta_{\alpha,\alpha'}\, { E\,+\varepsilon_\alpha\,+\,
{\Sigma_\alpha(E)\over 2\varepsilon_\alpha}\over 2\varepsilon_\alpha}
\,G_\alpha(E)\nonumber\\
G_{-\alpha^\dagger -\alpha'}(E) &=&\delta_{\alpha,\alpha'}\, { -E\,+\varepsilon_\alpha\,+\,
{\Sigma_\alpha(E)\over 2\varepsilon_\alpha}\over 2\varepsilon_\alpha}
\,G_\alpha(E)\nonumber\\
G_{-\alpha^\dagger {\alpha'}^\dagger}(E) &=& G_{\alpha -\alpha'}(E)= 
\delta_{\alpha,\alpha'}\, 
{\Sigma_\alpha(E)\over 4\varepsilon^2_\alpha}\, \,G_\alpha(E) ,
\end{eqnarray}
where the full propagator is: 
\begin{equation}
G_{\phi_\alpha \phi^\dagger_{\alpha'}}(E)=\delta_{\alpha,\alpha'}\,G_\alpha(E)=
\delta_{\alpha,\alpha'}\,\left(E^2\,-\varepsilon^2_\alpha\,-\Sigma_\alpha(E)\right)^{-1}.
\end{equation}
The mass operator being given by:
\begin{equation}
\Sigma_\alpha(E)={b^2\,s^2\over 2}{I_\alpha(E)\over 
1-{b\over 2}\,I_\alpha(E)}~.
\end{equation}

\smallskip\noindent
For what concerns the 2p-1h and 2p-2p Green's functions we introduces indices $i$ to label 
the destruction (creation) operators: $1=\beta, \beta' (\beta^\dagger, {\beta'}^\dagger)$, 
$2=(\beta, -{\beta'}^\dagger)_{sym} ((\beta^\dagger, -\beta')_{sym})$ and  
$3=-\beta^\dagger,  -{\beta'}^\dagger (-\beta, -\beta')$.
The results are: 
\begin{eqnarray}
G^{(i)}_{\beta\beta', \alpha^\dagger}(E) &=& 
G^{(i)}_{\alpha^\dagger ,\beta\beta'}(E)\nonumber\\
&=& {b\,s\over \sqrt V}\,\delta_{\alpha-\beta-\beta'}\,
{I^{(i)}_{\beta\beta'}(E)\over 1-{b\over 2}\,I_\alpha(E)}\,
\left({E\,+\,\varepsilon_\alpha\over 2\,\varepsilon_\alpha}\right)\,
G_{\phi_\alpha \phi^\dagger_\alpha}(E)\nonumber\\
G^{(i)}_{\beta\beta', -\alpha}(E) &=& 
G^{(i)}_{-\alpha ,\beta\beta'}(E)\nonumber\\
&=& {b\,s\over \sqrt V}\,\delta_{\alpha-\beta-\beta'}\,
{I^{(i)}_{\beta\beta'}(E)\over 1-{b\over 2}\,I_\alpha(E)}\,
\left({-E\,+\,\varepsilon_\alpha\over 2\,\varepsilon_\alpha}\right)\,
G_{\phi_\alpha \phi^\dagger_\alpha}(E)\label{A12}
\end{eqnarray}

\begin{eqnarray}
G^{(i j)}_{\beta\beta', \,\gamma\gamma'}(E) &=& 
I^{(i)}_{\beta\beta'}(E)\,\delta_{i,j}\,\left(\delta_{\beta\gamma}\,
\delta_{\beta'\gamma'}\,+\,\delta_{\beta\gamma'}\,\delta_{\beta'\gamma}\right)\nonumber\\
& &+\,{b\over V}\,\sum_{\alpha}\,
{\delta_{\alpha-\beta-\beta'}I^{(i)}_{\beta\beta'}(E)\quad
\delta_{\alpha-\gamma-\gamma'}I^{(j)}_{\gamma\gamma'}(E)\over 
1-{b\over 2}\,I_{\alpha}(E)}\nonumber\\
& &+\,{b^2\, s^2\over V}\,\sum_{\alpha}\,
{\delta_{\alpha-\beta-\beta'}I^{(i)}_{\beta\beta'}(E)\quad
\delta_{\alpha-\gamma-\gamma'}I^{(j)}_{\gamma\gamma'}(E)\over 
\left(1-{b\over 2}\,I_{\alpha}(E)\right)^2}\, 
G_{\phi_{\alpha} \phi^\dagger_{\alpha}}(E).
\end{eqnarray}

Finally, we give the useful relations:
\begin{align}
\sum_{i=1,2,3}\sum_{\beta\beta'} \frac{1}{\sqrt V} b\,s\, 
\left( G^{(i)}_{\beta\beta', \bar\alpha}(E) + G^{(i)}_{\beta\beta', -\alpha}(E) \right) &=
      2\, \Sigma_\alpha(E) \,
      G_{\phi_\alpha \phi^\dagger_\alpha}(E) \label{combi covariante G3}
\end{align}
and
\begin{eqnarray}
\sum_{i,j \in {1,2,3}}\sum_{\beta\beta'\gamma\gamma'} G^{(i j)}_{\beta\beta', \,\gamma\gamma'}(E) &=& 
2\, \sum_{\alpha} I_\alpha(E)
+ \,b \,\sum_{\alpha}\,
{ \left( I_{\alpha}(E) \right)^2
  \over 
  1-{b\over 2}\,I_{\alpha}(E)
} \nonumber \\
&& +\,b^2\, s^2\,\sum_{\alpha}\,
{ \left( I_{\alpha}(E) \right)^2
  \over 
  \left(1-{b\over 2}\,I_{\alpha}(E)\right)^2
}\, 
G_{\phi_{\alpha} \phi^\dagger_{\alpha}}(E)
\label{combi covariante G4}
\end{eqnarray}
These particular combinations are explicitly covariant if we forced it in the loop
$I(E,p)=I(E^2-p^2)$.

\end{document}